\documentclass[fleqn,10pt]{wlscirep}
\usepackage{cite}
\usepackage{amsmath,amssymb,amsfonts}
\usepackage{algorithmic}
\usepackage{graphicx}
\usepackage{textcomp}
\usepackage{cite}
\usepackage{graphicx}
\usepackage{enumitem}
\usepackage{textcomp}
\usepackage{mathtools}
\usepackage{subcaption}
\usepackage[utf8]{inputenc}
\usepackage[T1]{fontenc}
\usepackage{booktabs}
\usepackage[mathcal]{euscript}

\title{A GPS Alternative using Electrical Transmission Grids as Precision Timing Networks}

\author[1,*]{Stephen Robson}
\affil[1]{Cardiff University, Advanced High Voltage Engineering Research Centre, Cardiff, CF24 3AA, United Kingdom}

\affil[*]{robsons1@cardiff.ac.uk}

\begin{abstract}
It is widely recognised that over-reliance on GNSS (e.g GPS) for time synchronisation represents an acute threat to modern society, and a diversity of alternatives are required to mitigate the threat of an outage. This paper proposes a GNSS alternative using time dissemination over national scale transmission or distribution networks. The method utilises the same frequency bandwidth and coupling technology as established power line carrier technology in conjunction with modern chirp Spread Spectrum modulation. The basis of the method is the transmission of a time synchronised chirp from a central substation. During GNSS operation, all substations can estimate the time of flight by correlating the received chirp with a time-synchronised local copy. During GNSS outage, time sychronisation to the central substation is maintained by correcting for the precalculated time of flight. It is shown that recent advances in chirp spread spectrum allow for a computationally efficient algorithm with the capacity to compute hundreds of thousand of chirp correlations every second, improving the resolution of the system. ATP-EMTP simulations of the method on large transmission networks demonstrate  sub-$\mu$s timing accuracy even in the presence of low SNR and impulsive noise. An FPGA based prototype is developed and experimental testing also demonstrates sub-$\mu$s accuracy for time dissemination over a distance of 700 m. Averaging over time allows operation down to -20 dB, which could extend the range of the system to a national scale.
\end{abstract}
\begin{document}

\flushbottom
\maketitle
%
%
\thispagestyle{empty}

\section*{Introduction}

The infrastructure which underpins modern society is reliant on accurate time synchronisation. Such deep dependence is now prevalent in computer networks, electricity transmission grids, transportation, telecommunication and emergency services. The loss of an accurate time reference represents a high risk event and a known critical threat to societal security \cite{BLACKETT}. Radio based time dissemination methods are particularly vulnerable to spoofing and jamming attack vectors. A diversity of GNSS alternatives are required to mitigate the risk of an outage.

On electricity networks, the need for accurate and low latency time synchronisation is well established. A comprehensive study of timestamping requirements for distributuon networks emphasises the importance of both accuracy and latency\cite{IEEE1588_DIST}. Applications which demand hardware timestamping, such as Fast Distributed Generation (DG) disconnection, state estimation and Islanding in Microgrids require time accuracies in the order of 1 $\mu$s and latencies of 1-50 ms. Although these applications can be served by GNSS and Precision Time Protocol (PTP) algorithms such as IEEE-1588, the vulnerability of the grid to acute losses of these technologies is reported as a major risk electrical supplies.

Alternatives for widescale time dissemination over large areas already exist. Notably, the enhanced Long Range Navigation (eLoRAN) system is recognised as a potentially viable alternative to GNSS for both location and timing \cite{elorandemo1,elorandemo2,elorandemo3}. The system uses a number of beacons to transmit a signal and the receiver calculates its position based on the well known time difference of arrival approach in combination with corrections to account for slower than speed of light propagation speeds over land and sea. Studies have demonstrated 20 m location and $\pm$100 ns timing accuracy in its differential form \cite{eloran}, which meets stratum-1 requirements. However, as a wireless system, its vulnerability to spoofing and jamming is comparable to that of GNSS.


In this paper, the National Grid Transmission Network is proposed as a Precision Timing Network (PTN), forming the basis of a non-wireless candidate for a GNSS alternative. It is shown that the propagation characteristics of typical extra-high Voltage (EHV) transmission lines present a low-latency, low-attenuation transmission medium for low energy signals coupled into their aerial modes, a technology already established in Power Line Carrier (PLC) communications. Since the vast majority of signal energy remains closely coupled into the power line itself, this method of time dissemination can propagate over extremely large areas whilst remaining largely immune to wireless spoofing or jamming, thus providing a potentially valuable alternative vector should GNSS fail. 

The use of CSS has recently been proposed for time dissemination on distribution networks, with reported timing accuracy of 7.5 $\mu$s \cite{CSSTIME}, making it suitable for most smart grid applications. However, for larger national scale transmission networks, the Time of Flight (TOF) between transmitter and receiver must be considered and accounted for. Furthermore, the use of averaging to improve the accuracy is not considered. More recently, a CSS time of arrival estimator based on LoRa modulation is proposed \cite{ROBSONTIME}, which provides a computationally efficient way to perform a huge number correlations. This was further refined into a new PLC modulation scheme which is resilient to low SNR regimes and extreme multipath using the principle of averaging \cite{robsonchirplc}.  

This work extends the idea of time dissemination to national scale transmission networks. A key tenet of the proposed scheme is the established practice of Power Line Carrier (PLC), which has been used commercially on Transmission Lines for several decades \cite{PLCC_experience1,PLCC_experience2,PLCC_experience3}. Modal theory - developed largely as a means to explain the propagation of power line carrier signals - reveals that several low attenuation aerial modes support long range communication on double circuit transmission lines\cite{wedepohl1}. Analysis and experimental verification has reported signal attenuation over typical transmission lines of around 0.1 dB/km. Here, these channels are investigated as mediums for time dissemination signals.

In this work, we utilise the idea of chirp based time dissemination as part of a wider methodology for obtaining widescale time synchronisation across arbitrarily large transmission networks. The scheme requires TOF estimation and averaging during GNSS availability to allow a continuation of time synchronisation during GNSS outage, based only on the transmission of a repeated chirp beacon. It is shown through ATP-EMTP simulation and experimentation that sub-$\mu$s accuracy can be achieved. 

\section*{Transmission Networks as Precision Timing Network}  \label{sectiontran}
Seminal experimental work in the 1960's and 1970's verified the existence of low latency and low attenuation aerial modes of propagation on EHV transmission lines \cite{PLCC1,PLCC2,PLCC3,PLCC4,perz1964method}. These aerial modes are responsible for long distance communication in Power Line Carrier (PLC) systems, which has subsequently been used commercially for several decades in protective relaying applications on transmission networks. The central hypothesis of the present paper is that the aerial modes of transmission lines can be used as the basis of a PTN to disseminate a GPS timing signal. In other words, signals can be sent from a GNSS aligned transmitter to receivers on the network, and the arrival time of these signals, coupled with information on the TOF, can be used to reconstruct a precise timing estimator. Later, the use of CSS will be proposed to provide the basic requirements of the system, but first, this section will investigate the propagation modes of a typical transmission line structure.

\begin{figure}[]
\centering
\setkeys{Gin}{width=\linewidth}
    \begin{subfigure}[b]{0.51\columnwidth}
        \centering
        \includegraphics{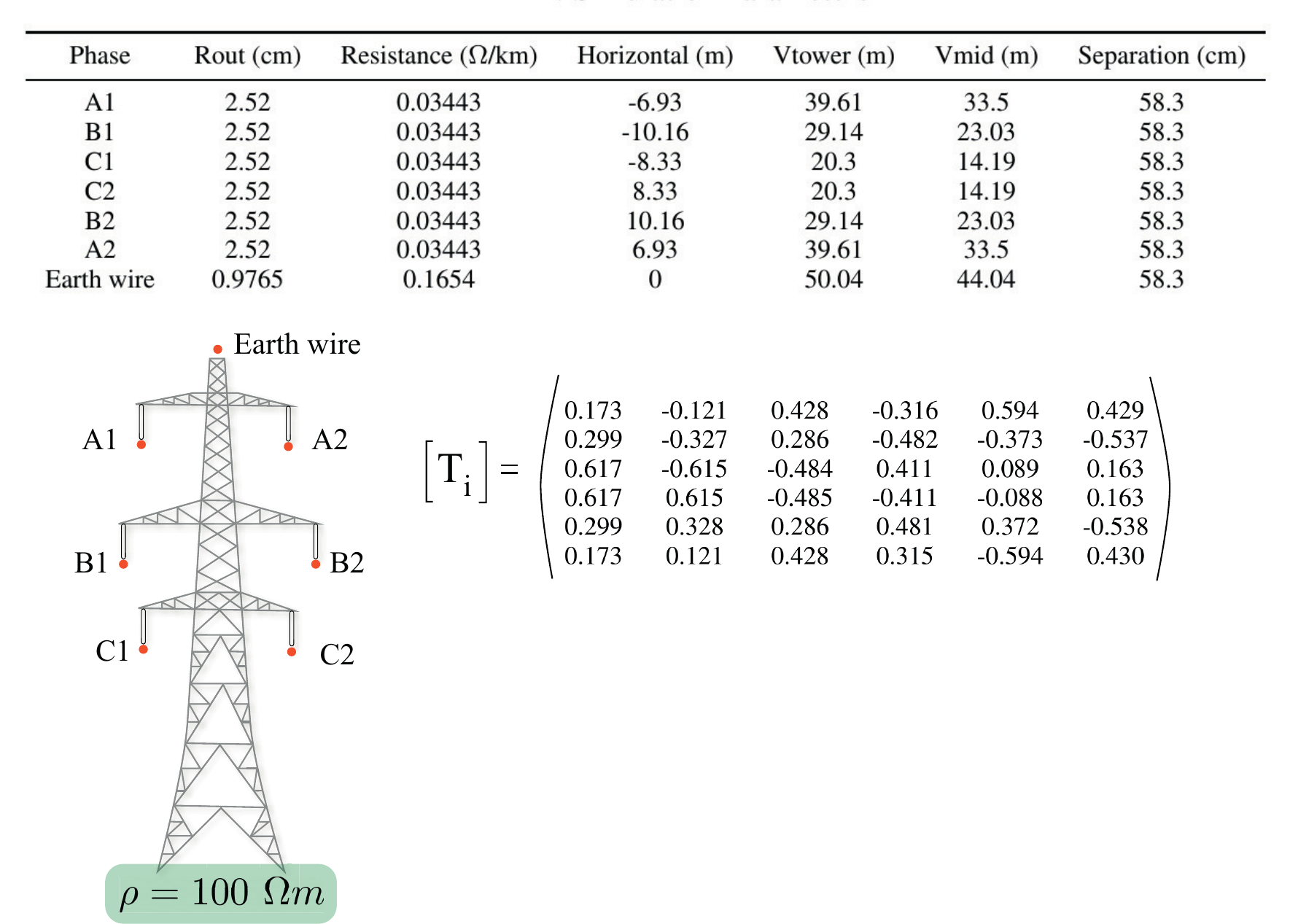}
        \caption{L6 Transmission Line}
        \label{fig:A}
    \end{subfigure}
    \begin{subfigure}[b]{0.24\columnwidth}
        \centering
        \includegraphics{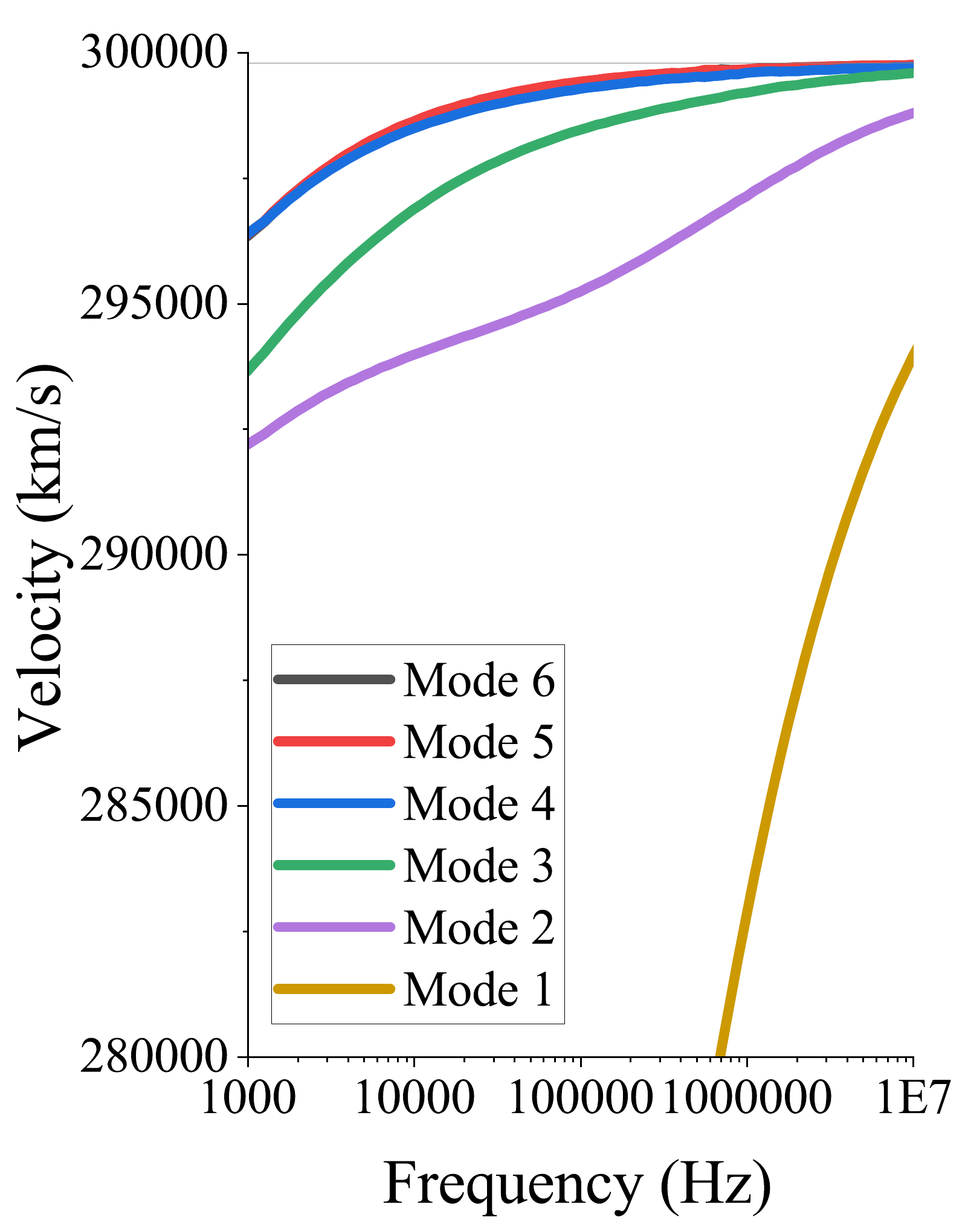}
        \caption{Velocity}
        \label{fig:vel}
    \end{subfigure}
    \hfill
    \begin{subfigure}[b]{0.24\columnwidth}
        \centering
        \includegraphics{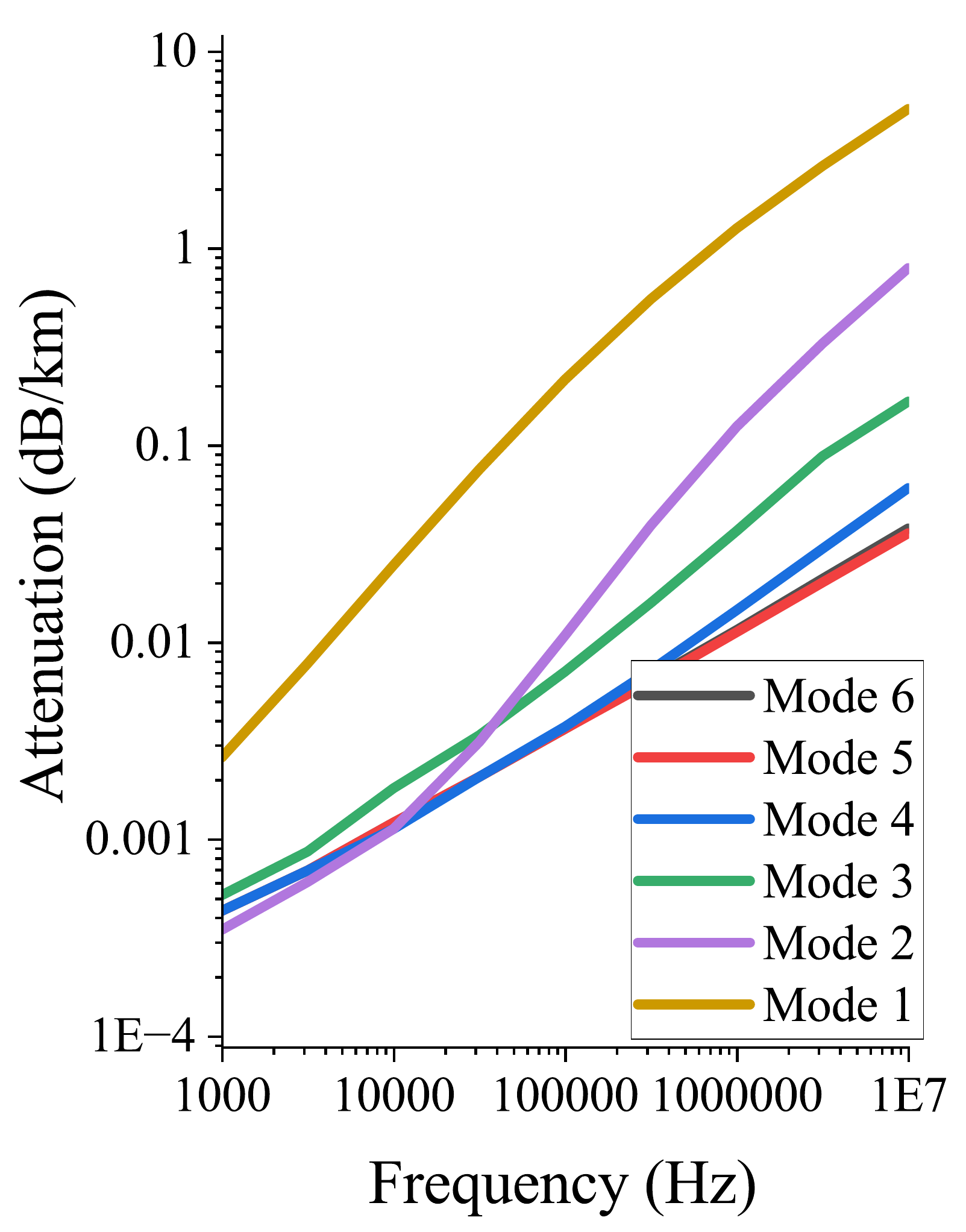}
        \caption{Attenuation}
        \label{fig:att}
    \end{subfigure}
		\caption{Velocity and attenuation of the L6 transmission line as a function of frequency and mode of propagation \label{line}}
\end{figure}

Fig.~\ref{line} shows details of the L6 transmission tower to be used in this study including its modal propagation parameters, expressed as the current transformation matrix, $[T_i]$, which has been calculated by the EMTP Lines and Cables Constants (LCC) routine. $[T_i]$ relates the phase domain and the modal domain according to $[I_{phase}] = [T_i] [I_{mode}] \label{mode_eqn1}$. It should be noted that the LCC uses a reduction technique to eliminate the earth wire, but this has no bearing on the subsequent analysis. In the modal domain, 6 independent modes exist. Fig.~\ref{fig:vel} and Fig.~\ref{fig:att} plots their respective velocities and attenuation as a function of frequency. The first major observation is the distinction between the singular `ground mode' and the 5 `aerial modes'. The ground mode (mode 1) uses the earth as the return conductor which leads to a significantly higher attenuation and slower velocity than the aerial modes, which use various arrangements of phase conductors for the forward and reverse path of the current. The difference in attenuation between the ground and aerial modes is approximately one order of magnitude which prevents the slower signal from causing ambiguity on the time of arrival estimate. It is shown later that small variations in the velocities of the aerial modes also have negligible impact on the timing estimator for distances typical of a national scale transmission system.

Observe that the modal propagation velocities of the aerial modes asymptotically approach close to the speed of light with little variation for frequencies above few 10's of kHz. Small variations in the conductor positioning, earth impedance or conductor resistivity have little impact on the velocity as it approaches its asymptote. Attenuation in the aerial modes (modes 2 to 6) are less than 0.1 dB/km at frequencies lower than 1 MHz, which confirms the early experimental work and practical experience of long range PLC.

The presence of low-latency and low-attenuation modes of propagation makes the transmission network a good candidate for a PTN, even over large geographical areas. The advantage of such a scheme is the possibility of widescale time dissemination in the event of a GNSS outage, and the invulnerability of the system to conventional jamming and spoofing techniques which are likely to disrupt wireless GNSS alternatives. In this work, we propose using CSS operating within similar bandwidths to traditional PLC schemes. The use of CSS allows accurate timestamping, advantageous propagation properties and the possibility of averaging over time to improve timestamping accuracy.


In this work, we propose a method of constructing a TOF estimate whilst whilst GNSS is operational and available to both the central node and the receivers (the calibration stage). The TOF - the delay between central node and receivers - can be continuously measured by each receiver if 1) The transmitter sends the beacon at exactly the top of the second, according to the rising edge of a 1PPS signal, and 2) Each receiver records the arrival of the beacon relative to the same 1PPS rising edge. Later, we propose the use of CSS and the use of statistical estimation to perform this calculation with continuous refinement over time, and the possibility to adjust for changes in propagation speed. The second step in the process commences the moment GNSS fails (the implementation stage). Now, each receiver loses its 1PPS time source, but the transmitter is able to switch to an accurate atomic clock and therefore continue to transmit a beacon which is aligned to the top of the second. Each receiver, with knowledge of the precise TOF between itself and the transmitter, can retain time synchronisation despite the loss of GNSS.

\section*{CSS Based Approach}

The proposed method requires a means to transmit a synchronised beacon from the central node to receivers. In this section, we propose the use of time-averaged CSS as a computationally efficient means of providing accurate timestamping, TOF measurement. First, the CSS approach utilised by the LoRa physical later is described.

\subsection*{Mathematical Description of the Proposed Method}
In recent years, the use of chirp waveforms for time dissemination on power networks has been proposed and investigated. Specifically, the method proposed in \cite{CSSTIME} and further explored in \cite{ROBSONTIME} is based on cross-correlating a received chirp, transmitted from a time-synchronized source through a distribution network, with a local, clean copy of the chirp. The chirp's excellent autocorrelation properties facilitate precise and repeatable timestamping. Recent advances have focused on how to make the process more computationally efficient to facilitate a practical implementation with the highest possible resolution. This development is important to the current method and will be addresses in this section.

 The autocorrelation function of a chirp is a sinc function with a well-defined peak corresponding to the moment of alignment. A raw implementation of this approach requires a sliding window correlation of the received signal with a local copy of the chirp. The idea to correlate a received chirp signal with a clean, local copy of itself is the core idea underpinning chirp based timestamping. However, this approach requires a full correlation to be performed for each new sample, which is computationally intensive. Since each new sample requires a full correlation, the resolution of the system is set by the sample time, and the computational complexity is, therefore, proportionate to the inverse of the resolution. In this work, the windowed approach used by LoRa to modulate information is used as the starting point. The windowed approach does not yield a new correlation result per sample, but instead yields a set of correlation results per window. The correlation peak, and in particular its offset, can be used to determine the precise time of arrival of the chirp. First, a brief description of the LoRa physical layer will be given.

Consider a chirp with $2^{SF}$ samples, where SF is the Spreading Factor. The symbol energy is $E_s$, $T$ is the LoRa sampling period, $B=\frac{1}{T}$ is the LoRa bandwidth and $k\in 0,1,2\dots 2^{SF-1}$ is an index representing the cyclic offset of the chirp. The $n=0,1,2\dots 2^{SF-1}$ sample digital symbol is given as:

\begin{equation}
 w_k(nT) = \sqrt{\frac{E_s}{2^{SF}}} e^{j2\pi \cdot (k+n) \mod 2^{SF} \cdot \frac{n}{2^{SF}}}  \label{eqn:2}
    \end{equation}

$w_k$ can be used to generate the so called base chirp ($i=k=0$), defined as the chirp with a cyclic offset of zero. Alongside the other chirps, it completes the set of basis functions across all possible values of $k$.

After transmission, $w_k$ is corrupted by the channel and noise, becoming $r_k$. Equation~\ref{awgncorr} shows the correlation process, expressed (in its raw mathematical form) as the inner product of the received symbol, $r_k$ and the complex conjugate of symbol $i$,  $w^*_i$. In the presence of Additive White Gaussian Noise (AWGN), $\phi_i$, the correlator output will yield a maximum of $\sqrt{E_s} + \phi_i$ when $i=k$, (assuming $k$ is the transmitted symbol) and $\phi_i$ elsewhere.

\begin{equation}
  y(i)=  \sum_{n=0}^{2^{SF}-1} r_k(nT)  \cdot  w^*_i(nT) = \begin{cases}
      \sqrt{E_s} + \phi_i & i=k\\
      \phi_i & i \neq k\\
    \end{cases}     \label{awgncorr}  
\end{equation}
		
Where	$w^*_i$ is the conjugate of all possible basis functions. When $i=k$ a correlation peak will occur and an arg max routine can be used to determine the index of the maximum of the correlator output, $y$. In LoRa, the emergence of a correlation peak reveals the transmitter communication symbol, which is chosen by the transmitter as one out of $k$ cyclically shifted chirps. The method has recently been adapted for Power Line Communications (LoRa-PLC) on networks exhibiting the dual problem of severe multipath interference and low Signal to Noise Ratios (SNR) \cite{robson_chirp}. Multipath effects are particularly pronounced on power networks and result in multiple, delayed copies of the incident chirp at the receiver. Interestingly, LoRa-PLC demodulates each of the delayed copies separately because they correlate strongly with the progressively time shifted bank of chirps making up the LoRa basis functions.  Plotting the correlation results in reverse order mimics the impulse response of the channel. In TOF calculations, the incident wave is of principal importance and later impulses can be disregarded provided they are far enough away. This will be explored in depth later on.
		

\subsubsection*{Improving the Computational Efficiency using the LoRa approach}
		
The approach described in Eqn.~\ref{awgncorr} requires $2^{SF}$ complex multiplication and adds for each of the $2^{SF}$ possible basis functions. A far more computationally efficient method is available based on the following. Consider the initial inner product between the received signal and the $2^{SF}$ basis functions:

\begin{equation}
  y(i)=  \sum_{n=0}^{2^{SF}-1} r_k(nT)  \cdot  \sqrt{\frac{E_s}{2^{SF}}} e^{-j2\pi \cdot (i+n) \mod 2^{SF} \cdot \frac{n}{2^{SF}}} =   \sum_{n=0}^{2^{SF}-1} r_k(nT)  e^{-j2\pi \cdot \frac{n^2}{2^{SF}}}\cdot  \sqrt{\frac{E_s}{2^{SF}}} e^{-j2\pi \cdot (i+n) \mod 2^{SF} \cdot \frac{n}{2^{SF}}}  \label{eqneff} 
\end{equation}

It is shown in [1] that the second exponential term in Eqn.~\ref{eqneff} is equivalent to $e^{-j2\pi q n \frac{1}{2^{SF}}}$ for all combinations of $n$ and $k$ meaning:

\begin{equation}
  y(i)=  \sum_{n=0}^{2^{SF}-1} r_k(nT)  e^{-j2\pi \cdot \frac{n^2}{2^{SF}}}\cdot  \sqrt{\frac{E_s}{2^{SF}}}e^{-j2\pi i n \frac{1}{2^{SF}}}  \label{eqneff3} 
\end{equation}

Eqn.~\ref{eqneff3} can be computed in two parts.  First, the $r_k(nT)  e^{-j2\pi \cdot \frac{n^2}{2^{SF}}}$ term is equivalent to the multiplication of the received signal, $r_k$ with a downchirp. This process of `dechirping' is described as:

\begin{equation}
d(nT)=   r_k(nT)  \overbrace{e^{-j2\pi \cdot \frac{n^2}{2^{SF}}}}^\textrm{downchirp}  \label{eqndechirp} 
\end{equation}

Leaving: 
\begin{equation}
  y(i)=  \sum_{n=0}^{2^{SF}-1} d(nT) \cdot  \sqrt{\frac{E_s}{2^{SF}}}e^{-j2\pi i n \frac{1}{2^{SF}}}  \label{eqnfft} 
\end{equation}

Eqn.~\ref{eqnfft} is equivalent to the Discrete Fourier Transform (DFT) operation on $d(nT)$, which is computed most efficiently as the FFT. It has been demonstrated that a single dechirp operation followed by a Fast Fourier Transform (FFT) is mathematically equivalent to performing the full set of correlations, but at a much lower computational complexity. This opens up the possibility of a huge number of correlations per window - and an associated increase of resolution - on standard devices like microcontrollers and FPGAs. 

In its basic form, the accuracy of this approach is limited to a resolution set by the distance between the bins in the correlation result, which in turn is set by the LoRa sampling period $T$ and, by extension, the LoRa bandwidth $B = \frac{1}{T}$. The resolution can be improved by decreasing $T$, which in turn increases $B$. The resolution can also be improved without the need to increase the bandwidth by repeating the demodulation process $s$ times with a progressive fine timing offset of $\delta = \frac{T}{s}$. For example, if the LoRa bandwidth is 100 kHz ($T= 10~ \mu s$), setting $s=100$ increases the resolution to 100 ns without requiring a bandwidth of 10 MHz. The process can be thought of as increasing the number of basis functions by a factor of $s$ with the distinction that the closely spaced chirps are not orthogonal, yet do have the property of revealing the sinc function and the moment of precise alignment. In its non-efficient form, the process is a correlation of the received signal, $r_k$, with $s \cdot 2^{SF}$ chirps which are spaced by $\frac{T}{s}$. $\mathbf{i} \in 0, 1, 2 \dots s \cdot (2^{{SF}}-1)$ is now the index of the correlation result with $s$ times as many entries and $\frac{T}{s}$ greater resolution than $i$.


\begin{equation}
  c(\mathbf{i})=  \sum_{n=0}^{s \cdot \left(2^{SF}-1\right )} r_k\left(n\frac{T}{s}\right)  \cdot  w^*_i\left(n\frac{T}{s}\right) 
     \label{longcorr}  
    \end{equation}

     \begin{figure}
        \centering
        \includegraphics[scale=0.33]{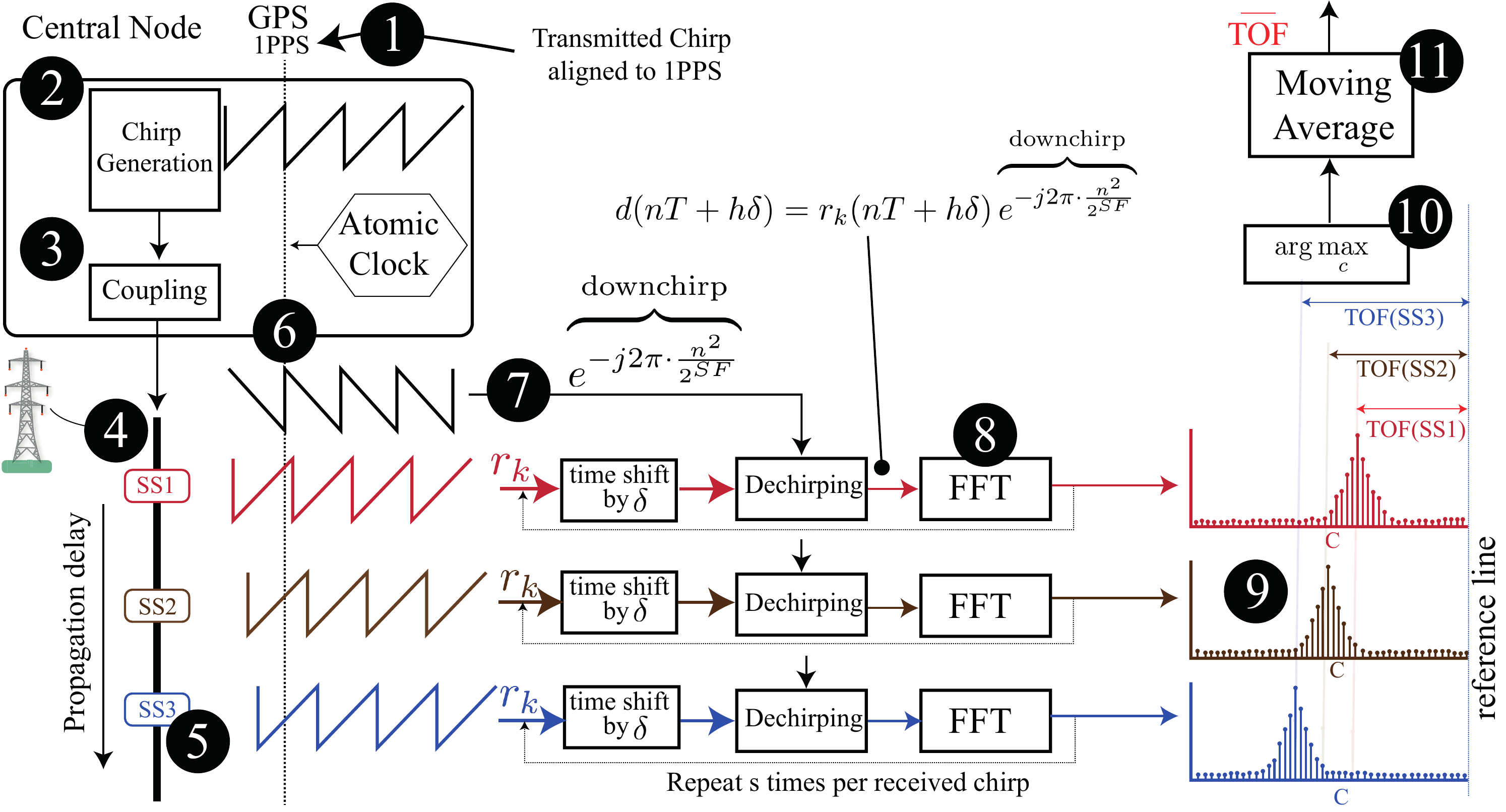}
        \caption{\textcircled{\raisebox{-0.9pt}{1}} The Central Node has access to a GNSS device, and a backup atomic clock, \textcircled{\raisebox{-0.9pt}{2}} A series of chirps are generated and aligned to 1PPS. \textcircled{\raisebox{-0.9pt}{3}} The chirp signal is coupled to the transmission line using established PLCC capacitive coupling. \textcircled{\raisebox{-0.9pt}{4}}The signal propagates through the network, predominantly on the aerial modes. \textcircled{\raisebox{-0.9pt}{5}} Each substation receives the chirps depending on the propagation delay. \textcircled{\raisebox{-0.9pt}{6}} In the calibration stage, the receiver also has access to GNSS 1PPS, and the received chirp has an offset proportionate to its propagation delay. \textcircled{\raisebox{-0.9pt}{7}} To guarantee this, the downchirp used in the dechirp process is aligned to \textcircled{\raisebox{-0.9pt}{8}} The FFT is performed, which reveals, \textcircled{\raisebox{-0.9pt}{9}}, $c$ correlation results (after $s$ fine tune operations). {\large \textcircled{\small 10}} The distance of the maximum correlation result from the reference line reveals the TOF. {\large \textcircled{\small 11}}  a moving average further improves the accuracy.    }\label{story}
    \end{figure}

Since it is not practical to compute Eqn.~\ref{longcorr} in hardware, the computationally efficient  approach splits the operation into $s$ smaller dechirp $\rightarrow$ FFT processes, each carried out with a time shift, as shown in Eqn.~\ref{aligned}. 

\begin{equation}
  c(i+h)=    \sum\limits_{n=0}^{2^{SF}-1} \overbrace{d(nT + h\delta)}^\textrm{GNSS Aligned} \cdot  \sqrt{\frac{E_s}{2^{SF}}}e^{-j2\pi i n \frac{1}{2^{SF}}}     \label{aligned}
  \end{equation}

	Where $h$ is an integer between 0 and $s$, representing the fine offset. The fine shift operation is applied to the received waveform, $r_k(nT + h \delta)$ (see Eqn.~\ref{eqndechirp}). An important aspect of this approach is ensure that the downchirp is GNSS aligned, meaning the first sample is concurrent with the rising edge of the 1PPS timing signal. The generation of the chirp at the central node (transmitter) is also GNSS aligned. This means the correlation result, $c$, will maximise at a distance from the rightmost index which is proportionate to the TOF of the chirp from the central node to the receiver. An overview of the proposed method is shown graphically in Fig.~\ref{story}. An FPGA implementation of this approach is shown later on.

\subsubsection*{Application of CSS to TOF measurement and Time Dissemination} \label{cssstory}
The essential observation in CSS based timestamping is that time delayed versions of the same chirp manifest in the demodulated correlation result at progressively lower positions. Delayed multipath energy reveals itself to the left of the reference line. This property is exploited in two ways to implement the proposed method. First, in the calibration stage, both the central node (transmitter) and receiver use local dechirp operations which are synchronised to a shared clock (e.g. GPS). The position of the right most spike in the correlation result will reveal the TOF based on its distance from the base chirp position (at index 0). Later spikes (further towards the left) can be disregarded for TOF calculations. Given that $c$ is the demodulation index, the maximum value of this index is defined as $\mathcal{D}  =   \arg \max_{c}$.

During the calibration stage, this equation yields the TOF as $\mathcal{D}  = TOF   =   \arg \max_{c} \mid \text{GNSS active}$, where $TOF$ is expressed in terms of the number of fine shift separations from the reference line, and, for simplicity, the reference line is designated the zero index, and the index becomes increasingly positive as it moves to the left. Conversion to time requires a multiplication by the time per fine shift, e.g. the resolution. Each demodulated chirp results in a new $TOF$ observation, and many such observations can be made within the calibration stage. To minimise the impact of noise, an estimator can be formed based on a moving average. Second, when GNSS is lost, the central node switches to an atomic clock and continues to send a time aligned chirp. The receivers, without a local source of time synchronisation, use the time of the received chirp, and previous knowledge of the TOF, to reconstruct a timing estimate.



\begin{figure}
        \centering
        \includegraphics[width=1\columnwidth]{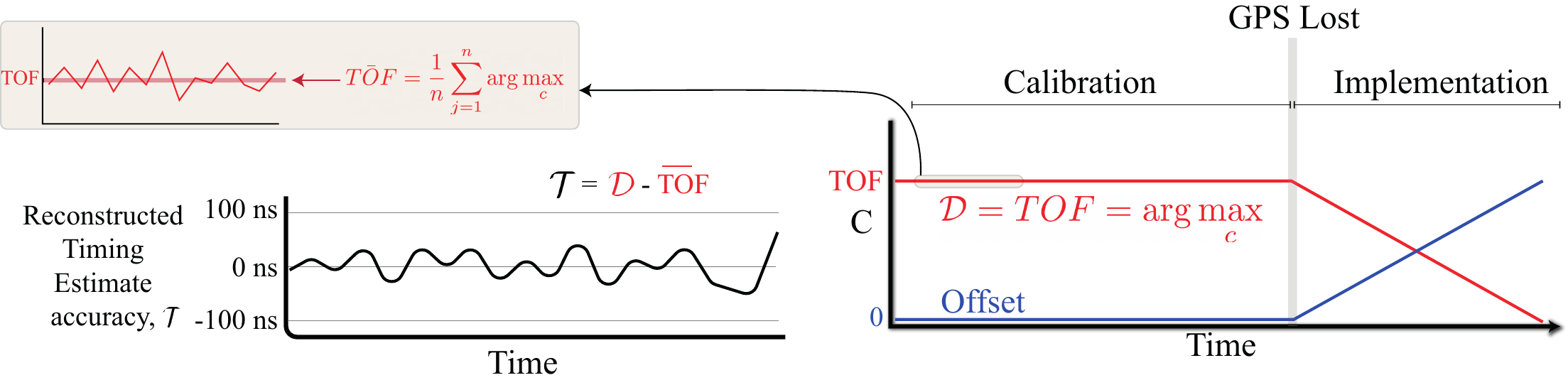}
        \caption{Evolution of the demodulation index during the transition between the calibration and implementation stages.} \label{full2}
    \end{figure}

Fig.~\ref{full2} shows the evolution of $\mathcal{D}$ in the transition between the calibration and implementation stages. In the calibration stage, $\mathcal{D} = TOF   =   \arg \max_{c}$ and is relatively constant (disregarding the error terms), and the moving average is used to build up a continuously improved $TOF$ estimate over time. The offset shows the difference between the 1PPS time sources at the central node and the receiver, which is close to zero during the calibration stage when both locations have access to GNSS. When GNSS is lost, the implementation stage begins and the offset becomes nonzero. The slope of the offset curve will depend on the drift of the local receiver clock relative to the central node's atomic clock.  As the receiver clock drifts away from the atomic clock of the central node, the offset increases. However, the demodulation index, $\mathcal{D}$, shifts as the exact inverse of this offset. Therefore, a timing offset, $\mathcal{T}$, can be formed as $\mathcal{T}  = \mathcal{D} - \bar{TOF}$. $\mathcal{T}$ in the implementation stage contains a correction for the $TOF$, and therefore provides full knowledge of the offset between the time of the receiver downchirp (used in the dechirp operation) and the time of the transmitted chirp. The disciplined 1PPS timepulse is therefore obtained by counting forward the remaining fraction of the second: $\text{1PPS}_{corrected}  = \text{1PPS}_{local} + (1-\mathcal{T})$. Note that this assumes a chirp of 1 second duration. In practice, each chirp duration will be 1-100 ms. A reconstructed 1PPS time pulse can still be reconstructed if an exact integer number of chirps fits within 1 second.



\section*{Simulation and Experimental Methods}
\subsection*{Noise Modelling}
Modelling the noise on an EHV or UHV line is challenging. Much theoretical and empirical work has been carried out to characterise the noise environment on low and medium voltage networks, revealing a mix of AWGN, coloured and impulsive noise \cite{fifytyyears}. At transmission level, corona noise must be added to this list \cite{corona,plcc_guide}. In this work, we propose the use of a general noise model based on the $\alpha$-stable distribution, S($\alpha$,$\beta$,$\gamma$,$\delta$)  \cite{alphastable1}. The Gaussian distribution is a special case of the  $\alpha$-stable distribution with the first shape parameter $\alpha=2$, the second shape parameter $\beta=0$, the scale parameter $\gamma =\frac{ \sigma}{\sqrt{2}} $ and the location parameter $\delta = \mu$, where $\sigma$ and $\mu$ are the standard deviation and mean of the resulting Gaussian distribution. However, by setting $\alpha<2$, the distribution exhibits heavier tails which model the impulsiveness of the power line channel.

Any communication scheme operating on the power line should be able to cope with high levels of impulsive noise against the general background of AWGN noise. A widely used mitigation technique is known as thresholding \cite{fifytyyears}. We propose the use of a thresholding technique based on clipping, similar to that recently utilised to improve the performance of Chirp-PLC in the presence of impulsive noise\cite{alphastable2,robsonchirplc}.  Below, it is shown that the $N_{th}$ noise sample, after thresholding,  is either the raw sample from the $\alpha$-stable distribution (if the value is less than the threshold, $T$) or $T$ itself if the same exceeds $T$.

\begin{equation}
    N_{th}=
    \begin{cases}
     N_{\alpha}(n), & \text{if}~ |N_{\alpha}(n)|<T \\
     T \cdot sgn(N_{\alpha}(n)), & \text{otherwise}
    \end{cases}
  \end{equation}

The threshold, $T$, is set to a user-defined multiple of the 90$^{th}$ percentile of the absolute value of the measured noise samples, $N_{90}$. Therefore, $ T = N_{90}\cdot M$, where $M$ is the multiple.  Since the $\alpha$-stable distribution does not exhibit finite variance the standard definition of SNR is invalid. Instead, the signal-to-dispersion ratio is used in this work, as expressed as  $SNR_D = 10\log_{10}\frac{P}{2\gamma^2}$, where $P$ is the signal power.



\subsection*{Simulation Methodology}
The simulation methodology utilises JMarti frequency dependent line models within the ATP-EMTP \cite{jmarti}, modelling the L6 transmission lines discussed earlier. The transmit chirp is generated within a parameterised C++ foreign model which is precompiled as a native model into the EMTP executable, with input arguments to determine the start frequency, end frequency and number of samples of the base chirp.  A consecutive series of chirps from this model is coupled via TACS into the transmission network via a coupling capacitive with properties similar to that used in power line carrier systems. The ATP-EMTP model is called via  DOS command from within the main Matlab script, meaning the entire methodology can be automated. Simulation results, particularly the signals from the substations are saved in the native .PL4 format and converted to the .MAT format using the PL42MAT software, which is called from within the Matlab script. The demodulation code is organised into objects, where each object executes the demodulation process with its own configurable set of parameters (bandwidth, spreading factor etc). Each object has methods for 1) Downconversion of the chirp to baseband, 2) Dechirping, 3) Running the FFT, and 4) Finding the index of the max result from the FFT. An inner loop introduces a progressive fine shift and repetition of the demodulation process $s$ times per window, as outlined earlier. Subsequent code pieces together the $s \cdot 2^{SF}$ correlation results into a single plot, and a final find max routine of this waveform is used to determine TOF estimate based on its distance from the reference chirp (e.g. the rightmost part of the waveform). The automated Matlab script allows Monte-Carlo simulations to be performed for arbitrary noise scenarios (varying SNR and impulsiveness, $\alpha$) across a user-defined number of runs. The Matlab code and EMTP models associated with this paper have been made available via GitHub.

 		\begin{table}[htbp]
  \centering
  \caption{Configuration Parameters of Experimental and Simulation Tests.}
    \begin{tabular}{lccc}
        \toprule
         Parameter                & Experimental & Simulation             \\
        \midrule
          Fclk (GPS Disciplined              & 10.485760 MHz   & 10 MHz                \\
          Slow Time Pulse (GPS disciplined)      & 1 Hz   & n/a          \\
						 Oversampling factor ($s$) & 32  & 100               \\
          Spreading Factor    & 10     & 13                \\
					Base Samples Per Chirp (2$^{SF}$)    & 1,024   & 8,192                  \\
						Oversampled Samples Per Chirp ($s \cdot$2$^{SF}$)    & 32,768 & 819,200                 \\
          LoRa Bandwidth  & 327.680 kHz  & 100 kHz                 \\
					Carrier Frequency & 1.25 MHz      & 170 kHz             \\
					Chirp Period & 3.125 ms          & 81.92 ms         \\
							Correlations per Chirp & 32,768   & 819,200               \\
       						Resolution & 95.367 ns   & 100 ns               \\
        \bottomrule
    \end{tabular}
  \label{tab:grav1}
\end{table}

\subsection*{Experimental Methodology}
\begin{figure*}
        \centering
        \includegraphics[width=1\columnwidth]{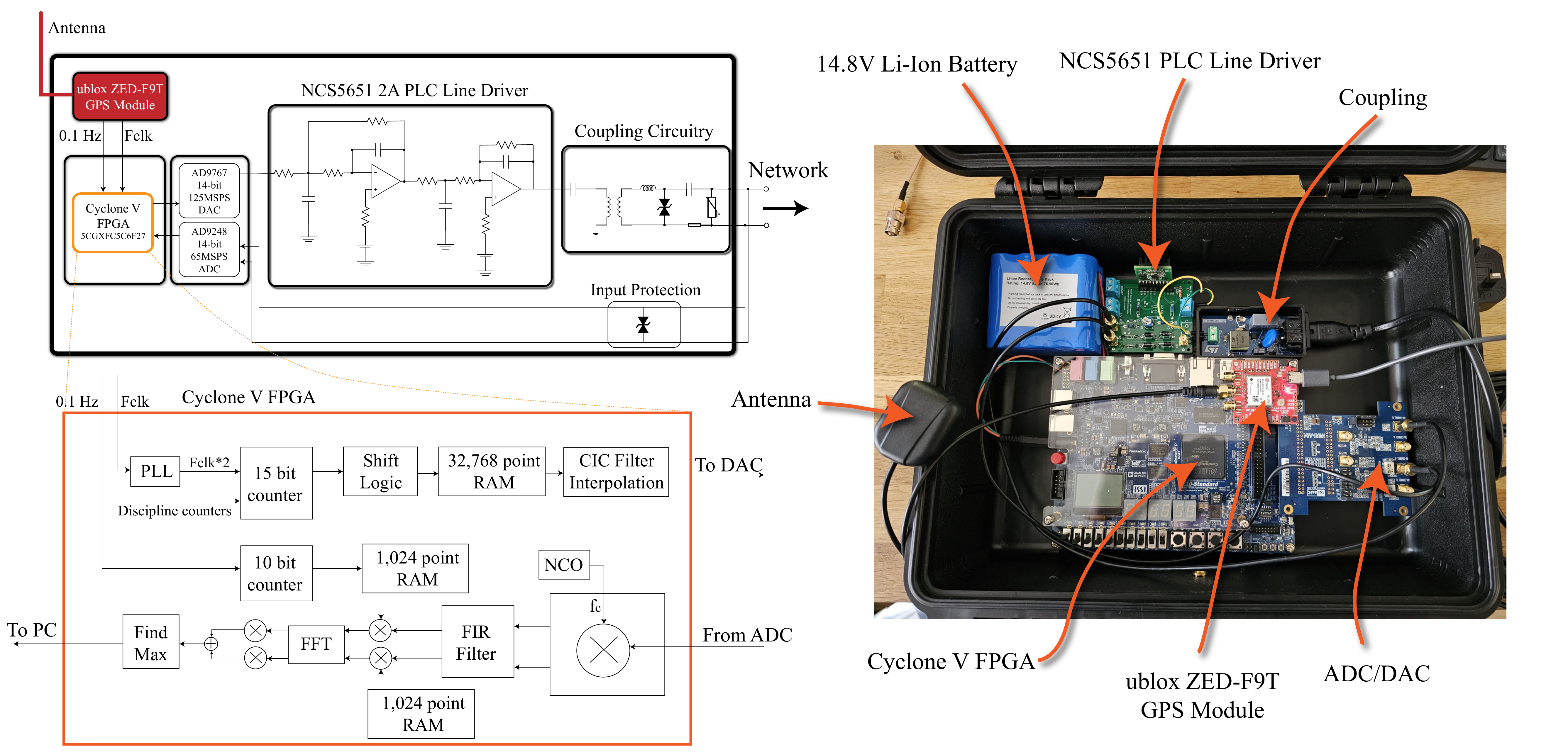}
        \caption{Prototype system     } \label{prototype}
        \label{fig:A}
    \end{figure*}

Figure~\ref{prototype} shows the architecture of the developed prototype. The system incorporates a u-blox ZED-F9T-00B timing module, which is a multi-band GNSS receiver with a 1-sigma timing accuracy of 5 ns. The module provides two GNSS disciplined timing signals, which is important to the developed technique (further explained later). The CSS based algorithm is implemented in FPGA hardware. The transmitter consists of a Phase Locked Loop which divides one of the GNSS disciplined timing signals by a factor of two. In the configuration shown, the fast timing signal from the timing module, Fclk, is set to 10.485760 MHz. The choice of frequency is to ensure that an integer number of 32 chirp waveforms fit within one 1 second cycle of the slower GNSS disciplined timing signal. A sequence of 32 progressively shifted chirps represents one full cycle from which the receiver can determine the accurate time of arrival to within a resolution of Fclk. A 15-bit counter operating at a rate of Fclk/reads samples from a pre-allocated 32,768 point RAM, which contains a full upconverted base chirp. The shift logic introduces a fine shift after every chirp. As noted previously, this is a computationally efficient way to increase the resolution of the system by a factor of $s$, where $s$ in this case is 32. A CIC filter provides interpolation prior to digital to analog conversion at a rate of 65 MSPS. The role of the receiver is to estimate the time of arrival of the chirp relative to the 1 Hz GNSS timing signal. Since each of the 32 chirps in a full cycle are progressively shifted, the correlation results will vary, reaching its peak when the shift happens to be closest in alignment to the receivers own chirp. Therefore, it is important that the receiver chirp is precisely aligned with the transmitter's base chirp. During the calibration stage, each receiver will have access to a GNSS clock, which guarantees this alignment. We define the Sclk - the LoRa bandwidth - with a clock which is exactly 32 times slower than Fclk/2. This slower clock is used to read from a pair of 10-bit RAMs to produce the real and imaginary parts of the baseband base chirp.

Tests are performed using a 700 m length of RG58 coaxial cable with a known propagation delay of $\approx 3.7 \mu s$. In the calibration stage, both the central node and the receiver have access to a separate GPS timing module with good access to the sky. Half way through the experiment, the receiver's antenna is disconnected and the 1PPS timing accuracy starts to drift.  To assess the accuracy of the timing estimate in the implementation stage, the offset between the transmitter's GPS 1PPS timepulse (the true time) and the unsynchronised receiver's 1PPS is compared. The index of the demodulation result should drift in proportion to this offset.

\section*{Results}


\subsubsection*{Effect of Mode Separation}

\begin{figure*}
\centering
\setkeys{Gin}{width=\linewidth}
    \begin{subfigure}[b]{0.32\linewidth}
        \centering
        \includegraphics{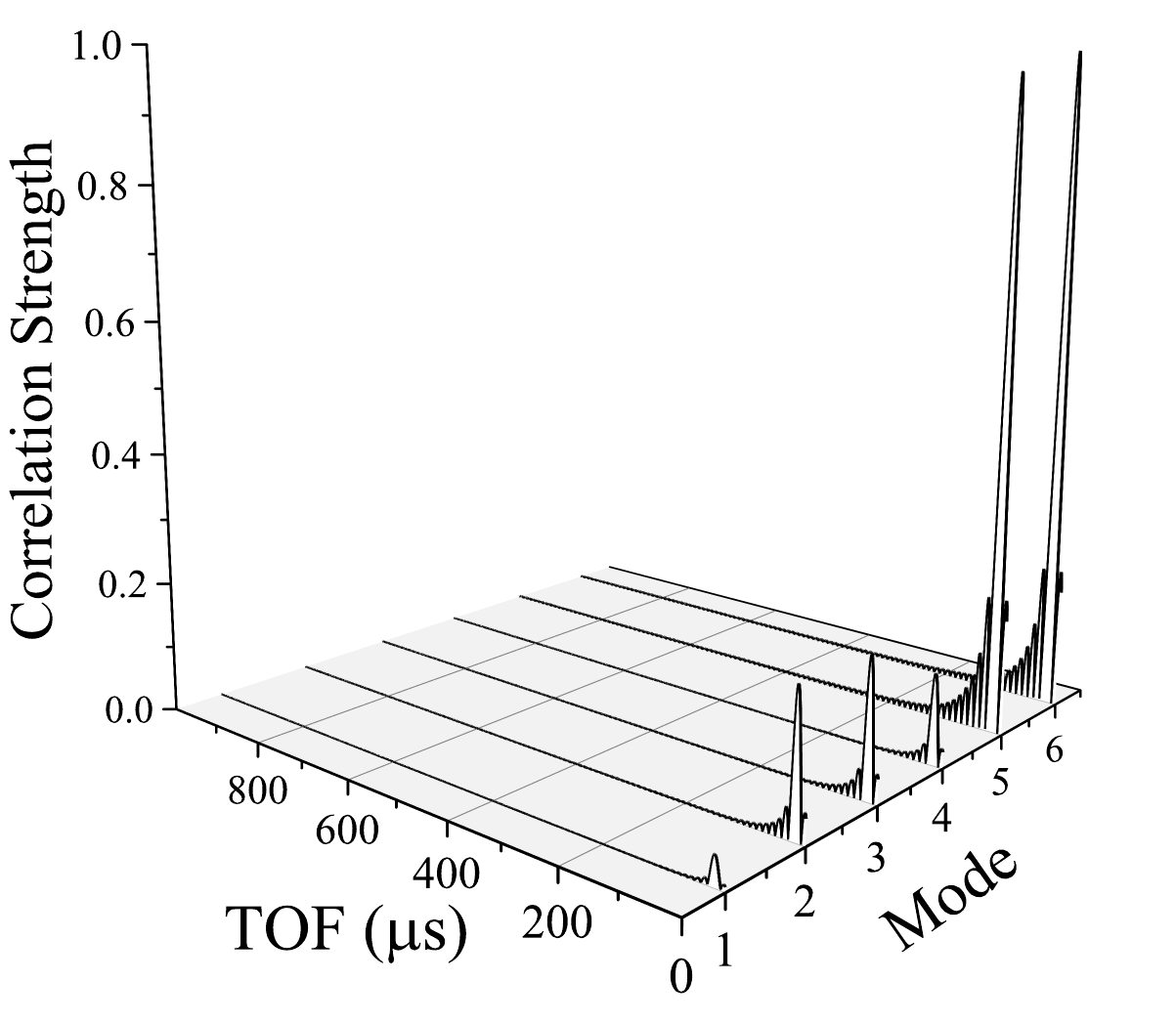}
        \caption{1 km}
        \label{fig:t1}
    \end{subfigure}
    \hfill
    \begin{subfigure}[b]{0.32\linewidth}
        \centering
        \includegraphics{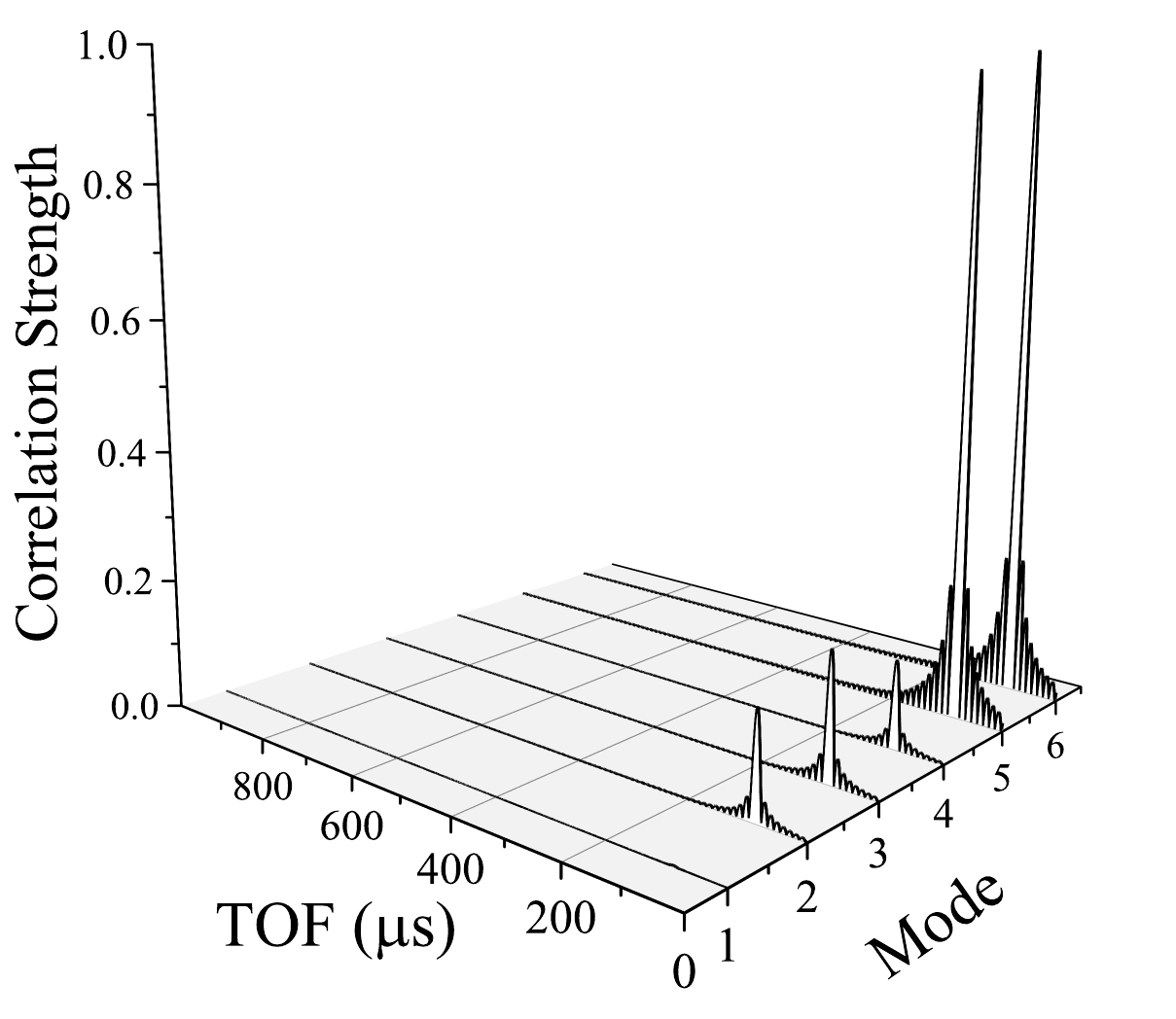}
        \caption{20 km}
        \label{fig:t2}
    \end{subfigure}
    \hfill
    \begin{subfigure}[b]{0.32\linewidth}
        \centering
        \includegraphics{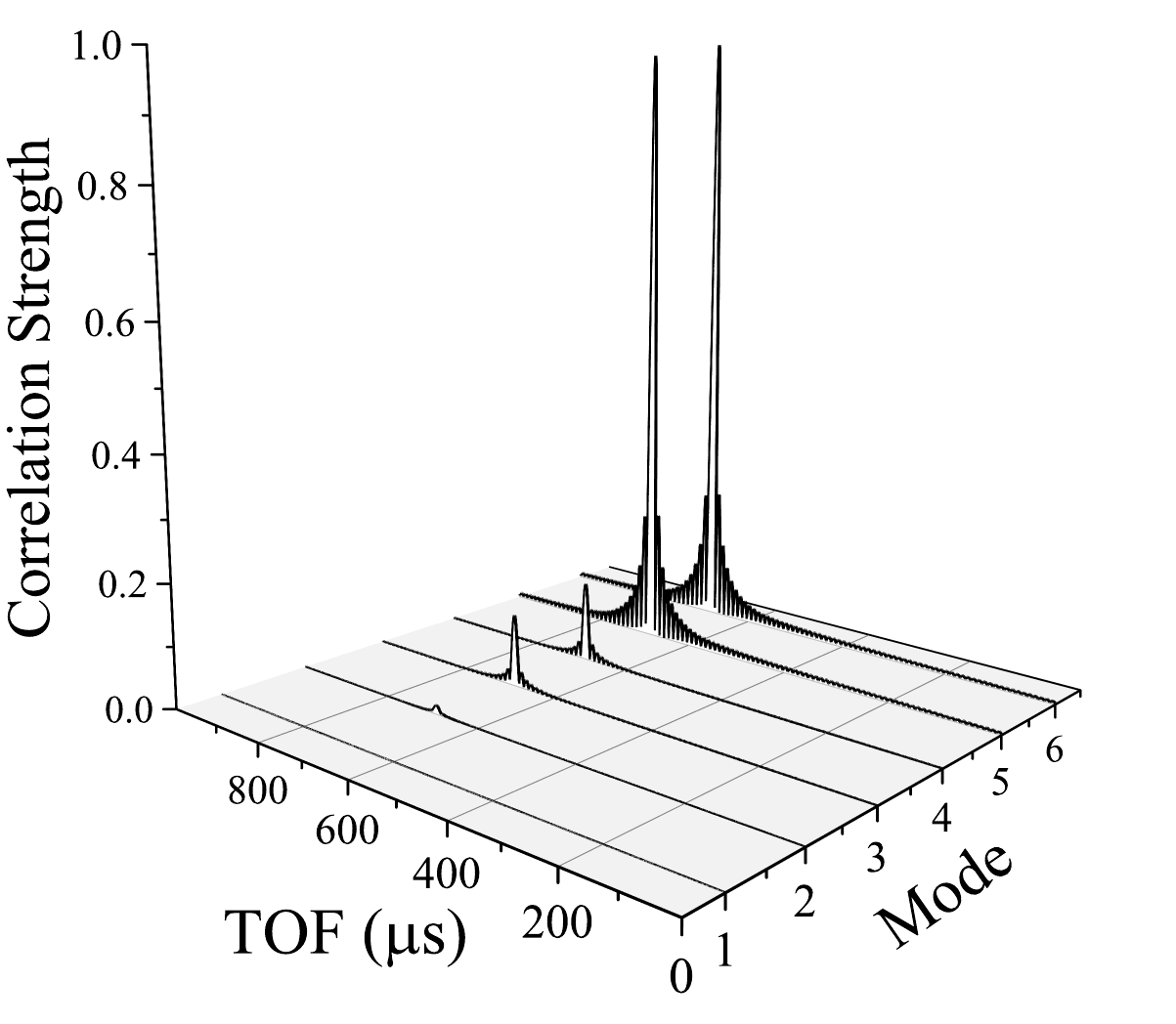}
        \caption{200 km}
        \label{fig:t3}
    \end{subfigure} 		
		\caption{Normalised correlation results ($c$) after modal decomposition of Phase A. The ground mode (1) quickly dissipates. Mode 2 is largely irrelevant at 100 km. At 200 km, aerial modes 3 to 6 remain with modes 5 and 6 dominating.\label{modes}}
\end{figure*}

The slight difference in velocities of the propagation modes could lead to ambiguity in the timing estimate. To investigate this, the received signal on Phase A, $[I_{phase}]$, is decomposed into its modal components,  $[I_{mode}]$, using $[I_{mode}] = [T_i]^{-1} [I_{phase}]$, where $[T_i]$ is the current transformation matrix as calculated by the EMTP LCC routine (and shown earlier in section~\ref{sectiontran}). Fig.~\ref{modes} shows the correlation results for all 6 modes at distances of 1 km, 20 km and 200 km, respectively. At 1 km (Fig.~\ref{fig:t1}), aerial modes 5 and 6 dominate but there are moderate contributions from the other 3 aerial modes (2, 3 and 4). The ground mode (1) makes a small contribution. At 10 km (Fig.~\ref{fig:t2}), the ground mode has largely dissipated. At 200 km (Fig.~\ref{fig:t3}), mode 2 is largely attenuated. As the distance increases, the signal is increasingly dominated by modes 5 and 6. It is important to note that the exact composition of modes will depend initially on the coupling arrangement, but as the signal propagates it is increasingly dominated by its two strongest aerial modes. This is an important result because the propagation speeds of modes 5 and 6 are similar, and the potential of dispersion due to the slower aerial modes is mitigated by their higher attenuations.


\subsubsection*{Performance of the Timing Estimate}
\begin{figure*}
        \centering
        \includegraphics[width=\columnwidth]{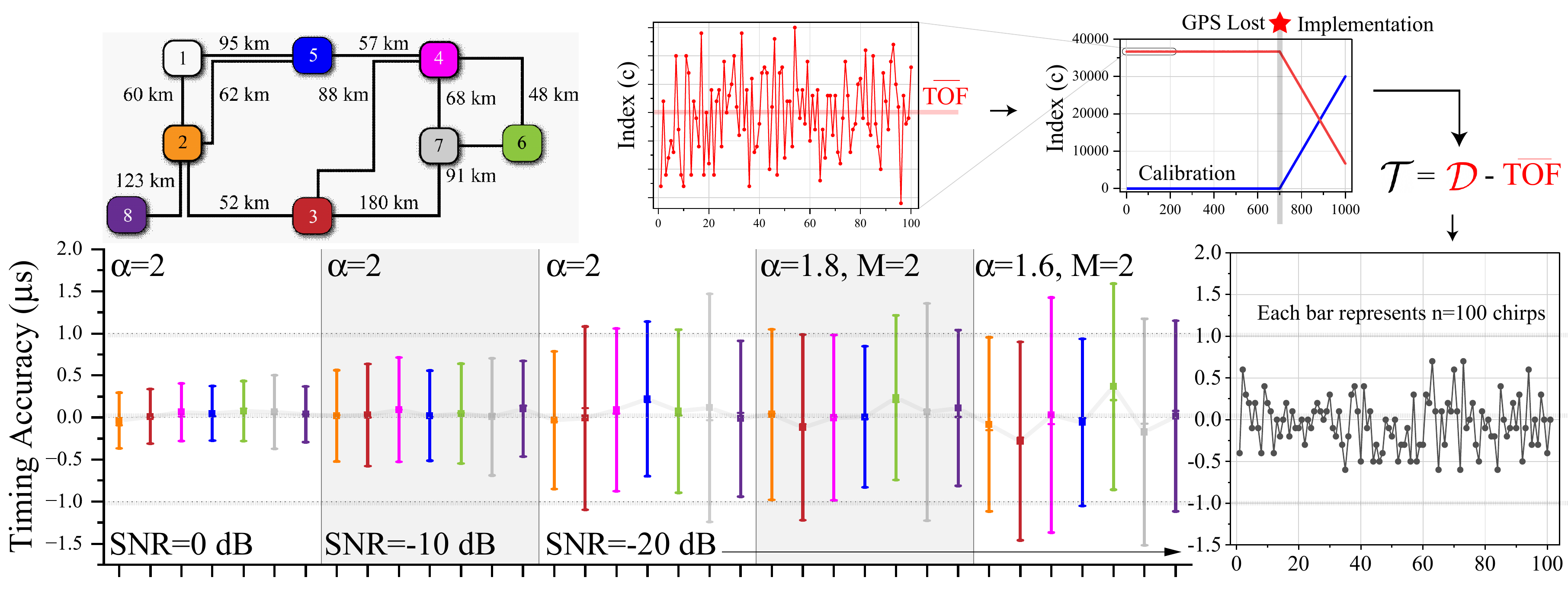}
        \caption{Simulated performance of the Timing Estimate for a calibration moving average of 900 chirps. The bars show n=100 chirp observations in the implementation stage. The boxplot's whiskers represent $\pm$ 1 standard deviation from the mean. } \label{performance}
        \label{fig:A}
    \end{figure*}

Figure~\ref{performance} shows the performance of the timing estimator for simulations performed on a large transmission network. We vary the SNR, degree of impulsiveness ($\alpha$) and the clipping threshold, M. 1000 Monte Carlo simulations are performed per condition, allocating 900 runs to the calibration stage and 100 runs to the implementation stage, where the latter experiences a progressive offset between the transmit and receiver clock phase. The 900 runs in the calibration stage are used to generate an estimate of the TOF ($\bar{TOF}$). A timing estimate is recovered in the implementation stage according to the method outlined previously.  We observe sub-$\mu$s timing accuracy for SNRs of -10dB and better. At SNR=0 dB, the standard deviation of the timing estimator is $\sigma = 250~ ns$. We also observe no deterioration in performance under progressively impulsive conditions ($\alpha=1.8, 1.6$).

\begin{figure*}
        \centering
        \includegraphics[width=\columnwidth]{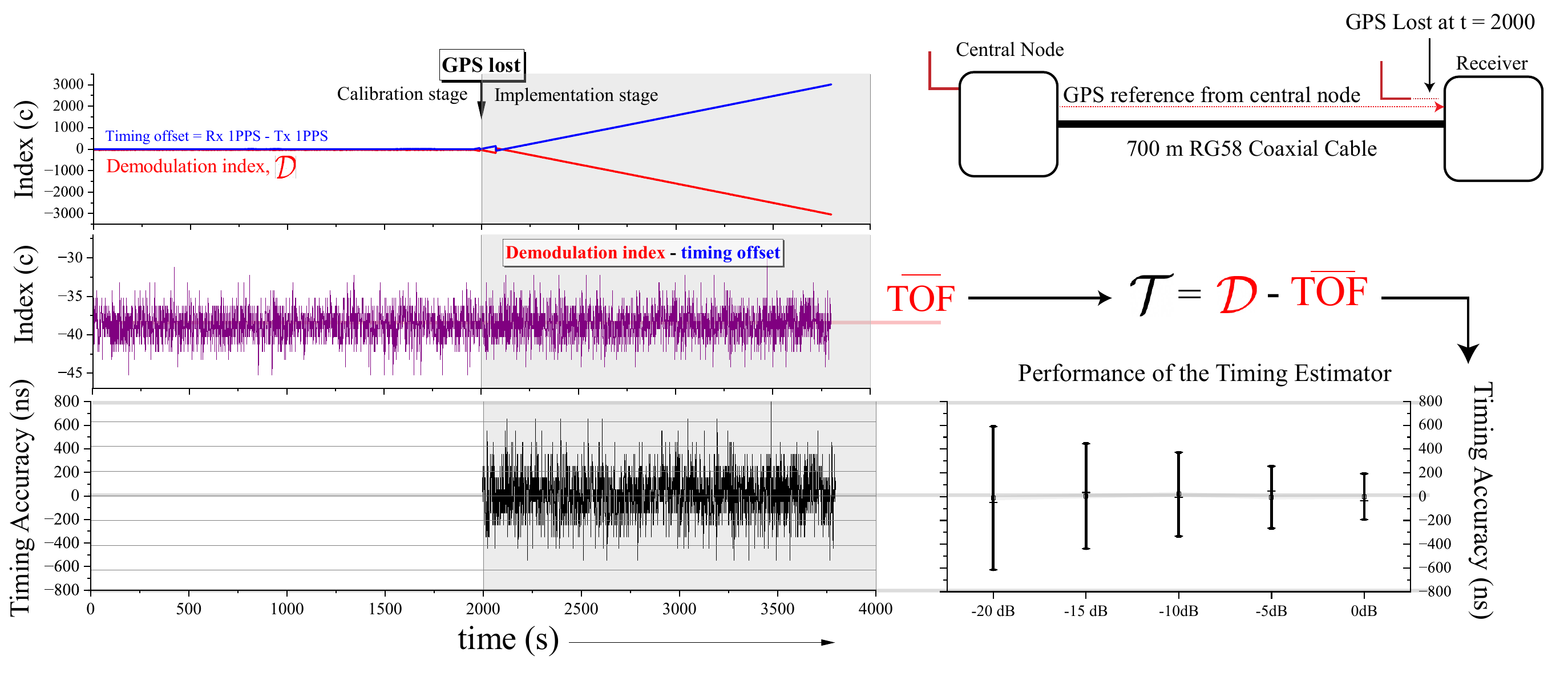}
        \caption{Experimental performance of the timing estimator for a calibration and implementation stage of 2,000 chirps, respectively. The boxplot's whiskers represent $\pm$ 1 standard deviation from the mean. } \label{performanceexp}
        \label{fig:A}
    \end{figure*}

Figure~\ref{performanceexp} shows the performance of the timing estimator for the experimental testing. Note that only AWGN is applied in this case, with SNRs ranging from 0 to -20 dB. Here, we perform calibration for 2,000 chirps and implementation for an additional 2,000 chirps, sampling a chirp once per second (total time of test of 33.33 minutes). We observe sub-$\mu$s timing accuracy for all cases.

\section*{Discussion}

The need for viable alternatives to GNSS is widely reported. The value of a national scale alternative which does not rely on wireless signals could provide a compelling backup to conventional satellite and other radio based timing services. This work has demonstrated the possibility of using a national scale transmission grid for time dissemination using a combination of established PLC technology and emerging CSS modulation techniques. We report sub-$\mu$s timing accuracy for performance on a simulated large transmission network and experimentally on a point to point coaxial cable of length 700 m. The use of CSS allows the method to operate at low SNRs - down to -20 dB, which opens up the possibility of widescale implementation over large geographical areas.

Although these results are promising, more work is required to address key limitations. An important assumption of the method is the existence of a relatively constant speed of propagation through the network. It is shown that the velocities of the aerial modes of propagation - responsible for long distance communication - asymptotically converge to the speed of light at frequencies above 100 kHz. However, weather variation and corona noise present a temporal set of variables which might change the propagation delays. Therefore, a key question is: how predictable and stable is the propagation velocity on transmission networks? In much the same way GPS corrects for atmospheric delay terms, the proposed approach could potentially address this issue with correction terms and predictive modelling, but further investigation is required.

\bibliography{master2}



\end{document}